\documentclass[tradiabstract]{aa} % for the abstract without %structuration
\usepackage{psfig}

\newcommand{\Lsun} {L$_\odot$}
\newcommand{\Msun} {M$_\odot$}

\begin{document}

%\thesaurus{02.16.2, 13.09.1, 11.09.4, 11.13.2, 11.09.1}

\title{Starburst and cirrus models for submillimeter galaxies}  

\author {A.~Efstathiou\inst{1}
	\and R.~Siebenmorgen\inst{2}
	}
\institute{
        School of Sciences, European University Cyprus, 
        Diogenes Street, Engomi, 1516 Nicosia, Cyprus.
\and 
 	European Southern Observatory, Karl-Schwarzschildstr. 2, 
	D-85748 Garching b.M\"unchen, Germany 
        }
\offprints{a.efstathiou@euc.ac.cy}
\date{Received October 21, 2008 / Accepted May 27, 2009}

\abstract{We present radiative transfer models for submillimeter
galaxies with spectroscopic redshifts and mid-infrared spectroscopy
from Spitzer/IRS and analyze available Spitzer/MIPS 24, 70 and
160$\mu$m data.  We use two types of starburst models, a cirrus
model and a model for the emission of an AGN torus in order to
investigate the nature of these objects. We find that for three of the
objects (25\%) cirrus emission alone can account for the
mid-infrared spectrum and the MIPS and submillimeter data.  For the
remaining objects we find that we need a combination of starburst and
cirrus in order to fit simultaneously the multi--wavelength data.
  We find that the typical submillimeter galaxy has comparable
luminosity in the starburst (median $L=10^{12.5} \ \rm{L}_\odot$) and
cirrus (median $L=10^{12.4} \ \rm{L}_\odot$) components. This could
arise if the galaxies have been forming stars continuously for the
last 250Myr with the star formation occurring in the last 5Myr being
shrouded by high-optical-depth molecular cloud dust, whereas the rest
of the starlight is attenuated by diffuse dust or cirrus with an
$A_V$ of about 1mag.  }

\titlerunning{RT models of submillimeter galaxies}

\maketitle

\section{Introduction}

The discovery of submillimeter galaxies (SMG) with the SCUBA
instrument mounted on the JCMT about a decade ago (Smail et al. 1997,
Hughes et al.  1998, Barger et al. 1998), and the realization that
most of them are at high redshift (Chapman et al. 2005, Dannerbauer et
al. 2004), was of particular significance for studies of galaxy
formation and evolution.  The implied high bolometric luminosities of
the SMGs prompted suggestions that they are distant analogs of local
ultra-luminous infrared galaxies (ULIRGs) which emit most of their
energy in the far-infrared part of the spectrum.

Over the last decade significant progress has been made in the
understanding of SMGs. Ivison et al. (2002) through deep radio mapping
of the areas covered by the 8mJy SCUBA survey (Scott et al. 2002) were
able to identify the radio and optical counterparts of a significant
fraction of the detected SCUBA sources. This allowed the determination
of photometric redshifts for the sources using the far-infrared/radio
correlation and other methods and the first estimates of their
luminosities and star formation rates which are found to be of the
order of 1000$\ M_\odot/yr$. Chapman et al. (2005) were able to obtain
spectroscopic redshifts for the radio identified sources and determine
a median redshift of 2.4. Alexander et al. (2005) studied the X-ray
properties of SMGs and concluded that the majority host an AGN which,
however, is not luminous enough to dominate the bolometric luminosity.

The diagnostic power of mid-infrared spectroscopy was first recognized
by the pioneering ground-based work of Roche et al. (1991). These
studies showed that starburst galaxies display emission features that
are attributed to PAH molecules whereas these features were absent
from the spectra of active galactic nuclei.  {\em ISO} studies
(e.g. Genzel et al. 1998) used mid-infrared spectroscopy to show that
ULIRGs are primarily powered by star formation. More recently {\em
Spitzer } studies extended this work to large samples of infrared
galaxies (e.g. Hao et al. 2007, Spoon et al. 2007). The unprecedented
sensitivity of the IRS on-board {\it Spitzer} made mid-infrared
spectroscopy of SMGs possible (Men\'{e}ndez-Delmestre et al. 2007,
Valiante et al. 2007, Pope et al.  2008). The spectra show that SMGs
display band emission of polycyclic aromatic hydrocarbon (PAH)
molecules and have similar mid-infrared spectra as the local starburst
galaxy M82.

In the local Universe, the infrared emission of normal star-forming
galaxies can be understood in terms of two components (Rowan-Robinson
\& Crawford 1989): starburst emission which is associated with
optically thick giant molecular clouds illuminated by recently formed
stars and {\it cirrus} emission which is associated with diffuse and
cold dust ($ T < 30$K) illuminated by the interstellar radiation
field. At a time when spectroscopic redshifts for SMGs were scarce,
Efstathiou \& Rowan-Robinson (2003; hereafter ERR03) showed that good
fits to the spectral energy distributions (SED) of SMGs with radio or
millimeter detections could be obtained with pure cirrus models or a
combination of cirrus and starburst with the submillimeter dominated
by cirrus.  The high $z$ cirrus models of ERR03
assumed an optical depth and intensity which is a factor of 2-3 higher
than the values found in local cirrus galaxies and the interstellar
radiation field was assumed to be that of a 250Myr old burst of star
formation.

Over the last two decades we have developed fairly sophisticated
radiative transfer models for AGN torus (Efstathiou \& Rowan-Robinson
1990, 1995), starburst (Rowan-Robinson \& Efstathiou 1993, Kr\"{u}gel
\& Siebenmorgen 1994, Efstathiou et al. 2000, Siebenmorgen \&
Kr\"{u}gel 2007) and cirrus (Siebenmorgen \& Kr\"{u}gel 1992,
Efstathiou \& Rowan-Robinson 2003). The starburst and cirrus models
take into account the effect of small grains and PAH
molecules. Starburst models have also been developed by Silva et
al. (1998), Takagi et al. (2003) and Dopita et al.  (2005). Our cirrus
model does not take into account the distribution of dust and stars
as  the GRASIL (Silva et al. 1998) or the Piovan et al. (2006) models
do  but it has been shown by ERR03 to give spectra that are in
good agreement with the spectral energy distributions of local cirrus
galaxies. As we show in section 3, the model is also in very good agreement
with observations of cirrus in our own galaxy. Other work on
radiative transfer modeling of galaxies has been presented by
Bianchi et al (1996), Xilouris et al. (1999) and Popescu et al (2000).
Other work on radiative transfer modeling of the torus in AGN
has been presented by Pier \& Krolik (1992), Granato \& Danese (1994),
Nenkova et al. (2002, 2008), Dullemond \& van Bemmel (2005), 
H\"{o}nig et al. (2006) and Schartmann et al. (2008).

In this paper we use radiative transfer models of starburst, cirrus
and AGN torus emission to constrain the properties of SMGs with
spectroscopic redshifts and mid-infrared spectroscopy and far-infrared
photometry (Sect.~2) from {\it Spitzer}.  For comparison two different
starburst models are applied: an {\it evolutionary} model 
(Efstathiou et al. 2000) that
incorporates a stellar population synthesis model and therefore
provides information about the stellar population that powers the
starburst (Sect.~4) and a {\it hot spot} model (Siebenmorgen \&
Kr\"{u}gel 2007) that is sensitive to the geometry of the dust
and stars and therefore provides information about the size of 
the starburst region (Sect.~5).  A flat Universe is
assumed with $\Lambda = 0.73$ and H$_0=71$km/s/Mpc.

\section{Observations}

Our sample is determined solely by the requirement that the galaxies
have been detected in the submillimeter and have mid-infrared
spectroscopy from the infrared spectrograph (IRS, Houck et al. 2004)
of the Spitzer Space Telescope (SST, Werner et al. 2004).  We model
all the galaxies in the PAH sample of Men\'{e}dez-Delmestre et
al. (2006), except SMM~J221733+001120 which is at low redshift and
which is clearly dominated by cirrus emission, and the objects in the
sample of Valiante et al. (2007).  The data in the sample of
Men\'{e}dez-Delmestre et al. cover the silicate feature and therefore
provide information about the extinction in the galaxies but the
objects in the sample of Valiante et al., being at higher redshift, do
not cover completely the silicate band. In total there are 12 objects
spanning a redshift range from 1.2 to 3.4 (Table 1).
Besides IRS, ISOCAM (Webb et al. 2003) and submillimeter photometry
(Smail et al., 2002, Scott et al., 2006, Ivison et al. 2005, Zemcov et
al. 2007) we complement the SED by retrieving all publicly available
data of the Spitzer far infrared imager (MIPS, Rieke et al. 2004) of
the targets in the three photometric channels centered at 24, 70 and
160$\mu$m.

\begin{figure*}
\vspace{-0.0cm}
\hbox{\hspace{-0.15cm}
\psfig{figure=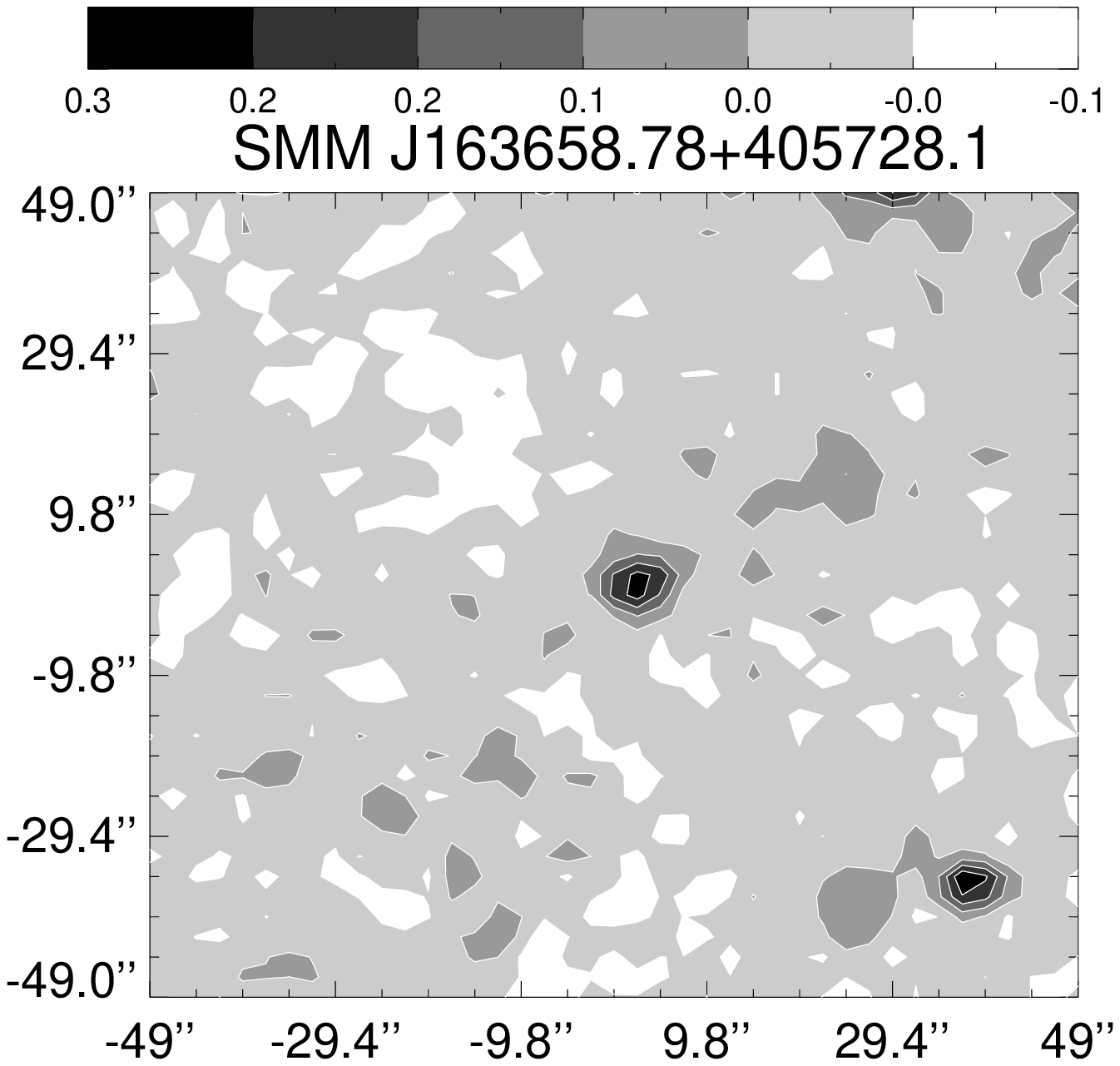,width=6.5cm}
\hspace{-0.7cm}
\psfig{figure=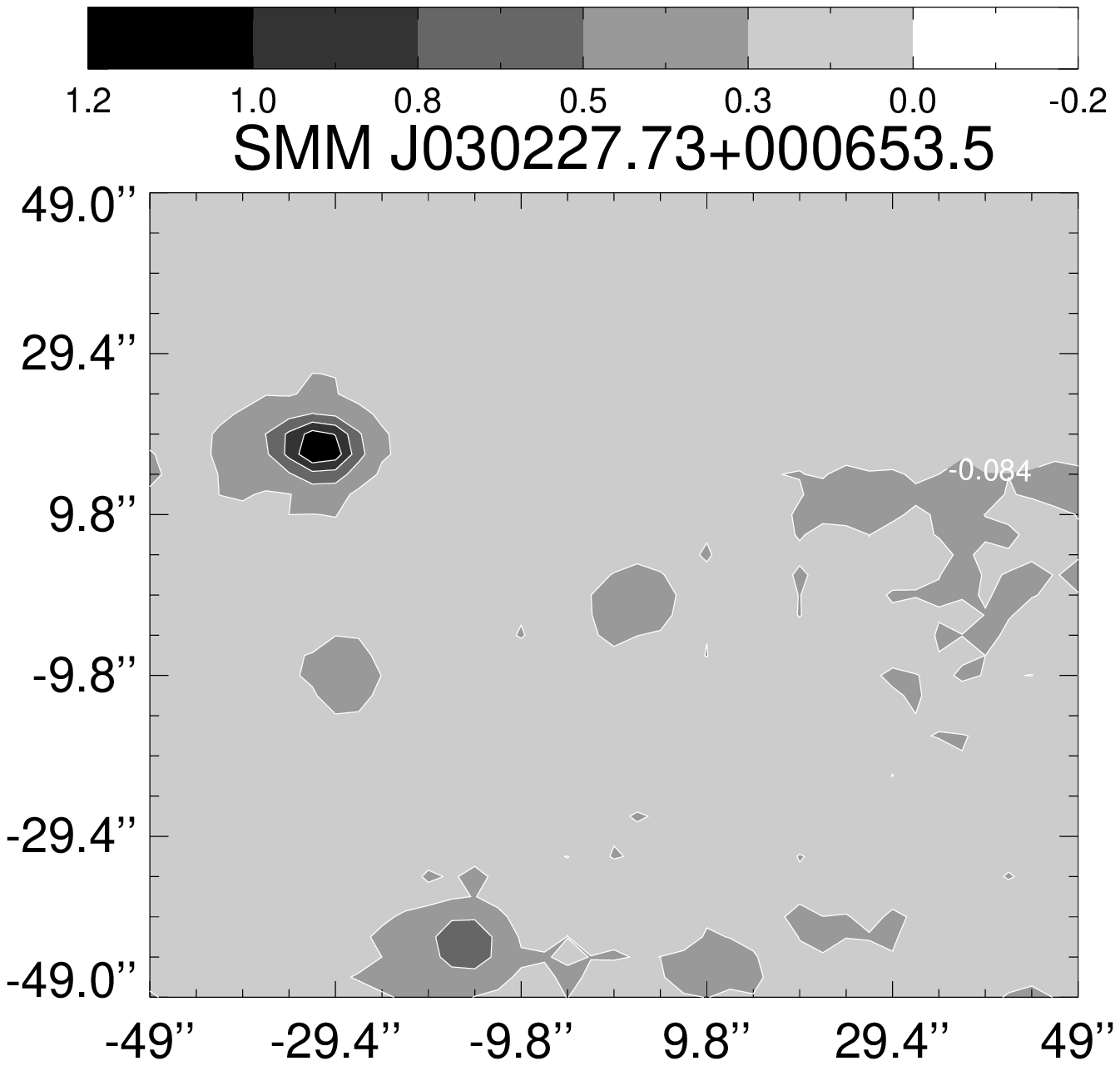,width=6.5cm}
\hspace{-0.7cm}
\psfig{figure=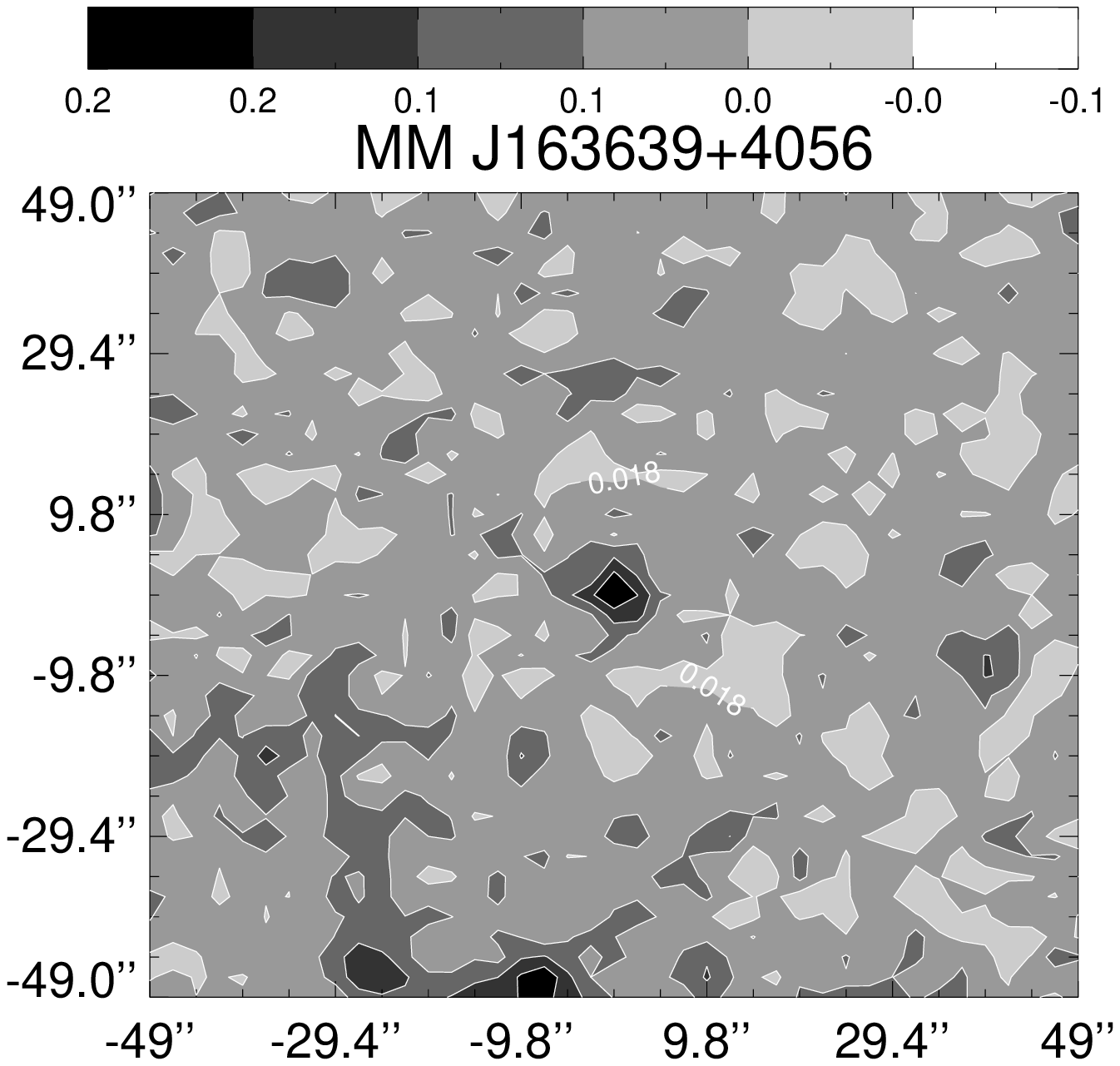,width=6.5cm}}
\vspace{-0.7cm}

\hbox{\hspace{-0.25cm}
\psfig{figure=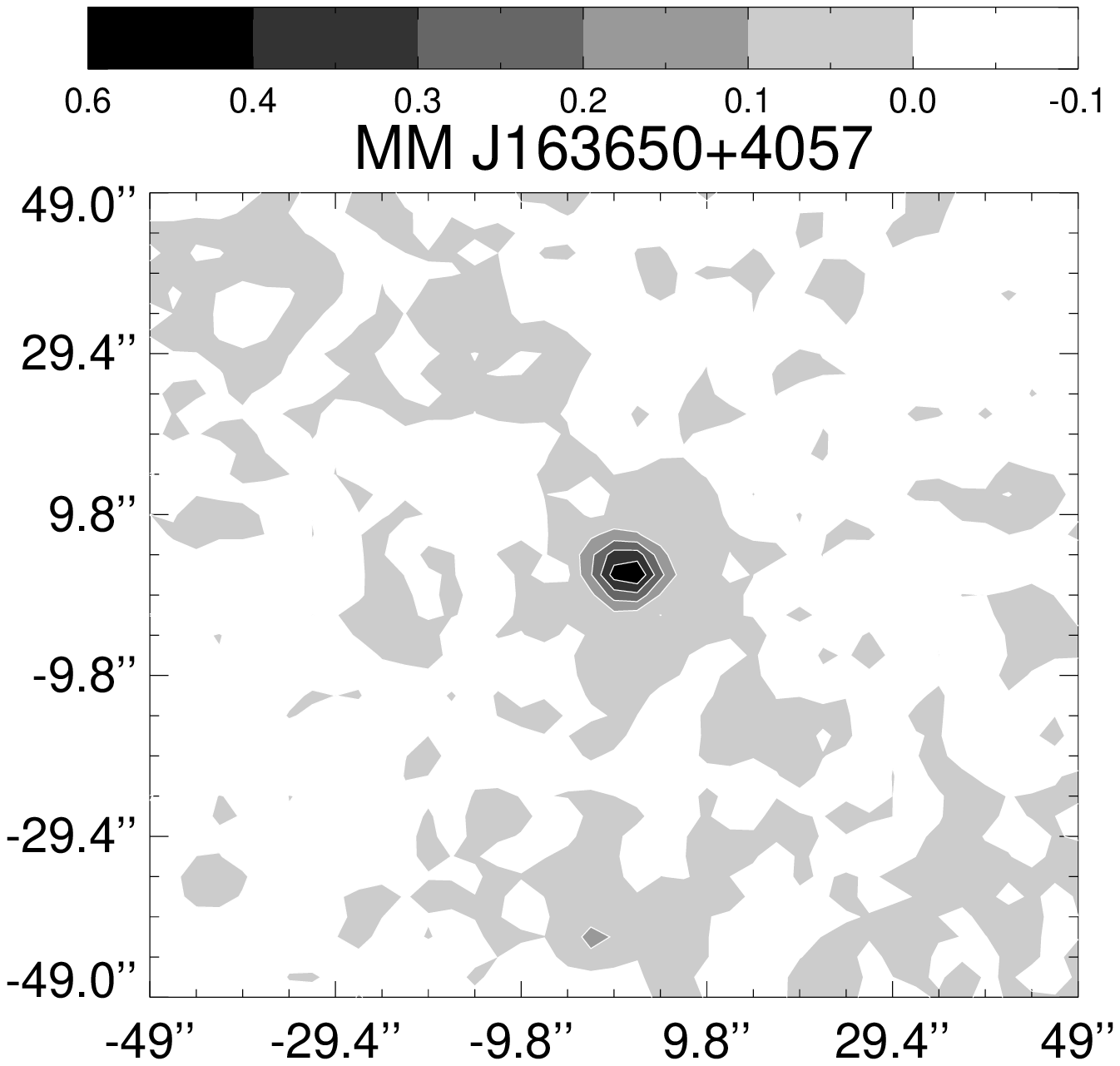,width=6.5cm}
\hspace{-0.7cm}
\psfig{figure=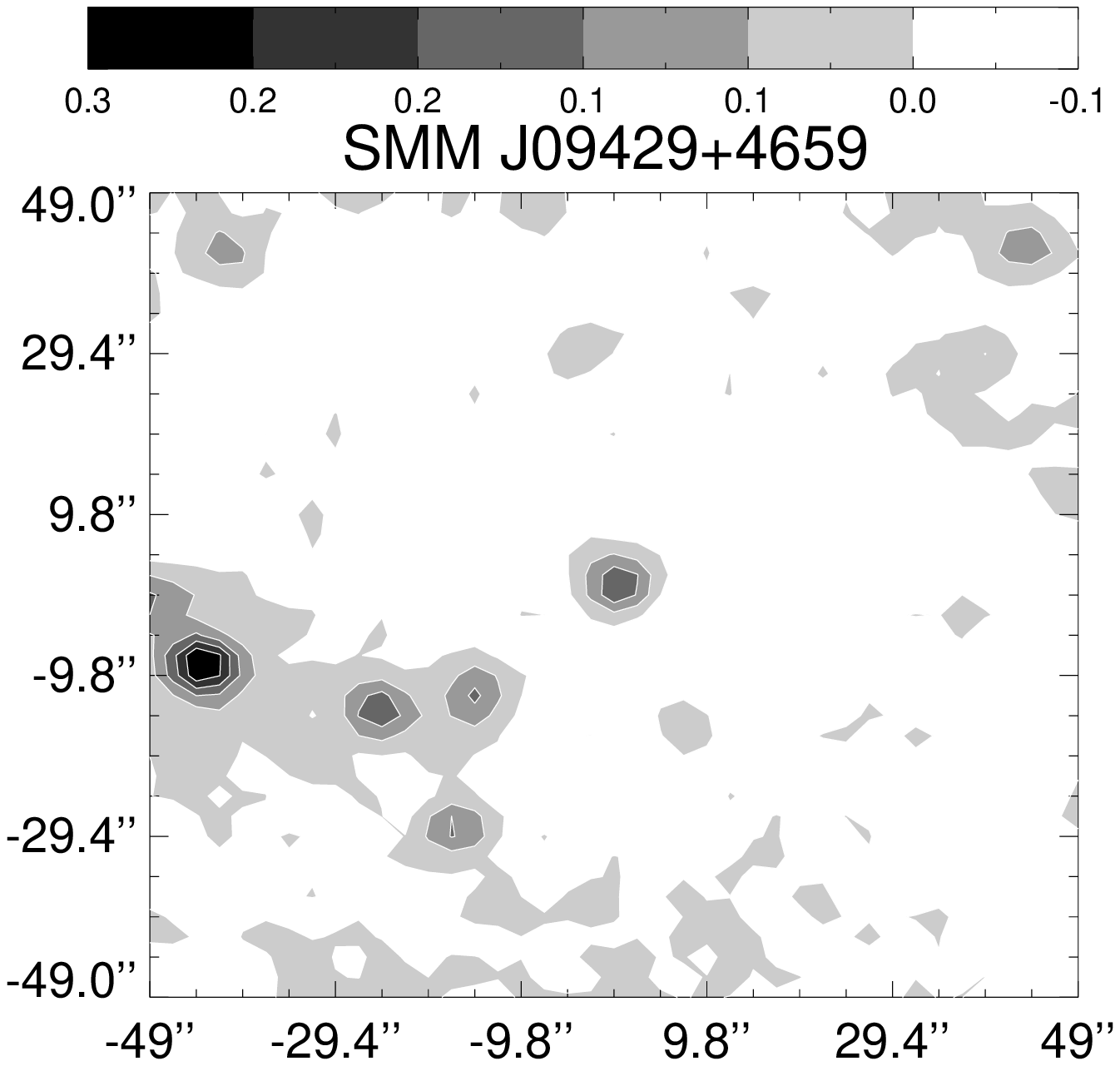,width=6.5cm}
\hspace{-0.7cm}
\psfig{figure=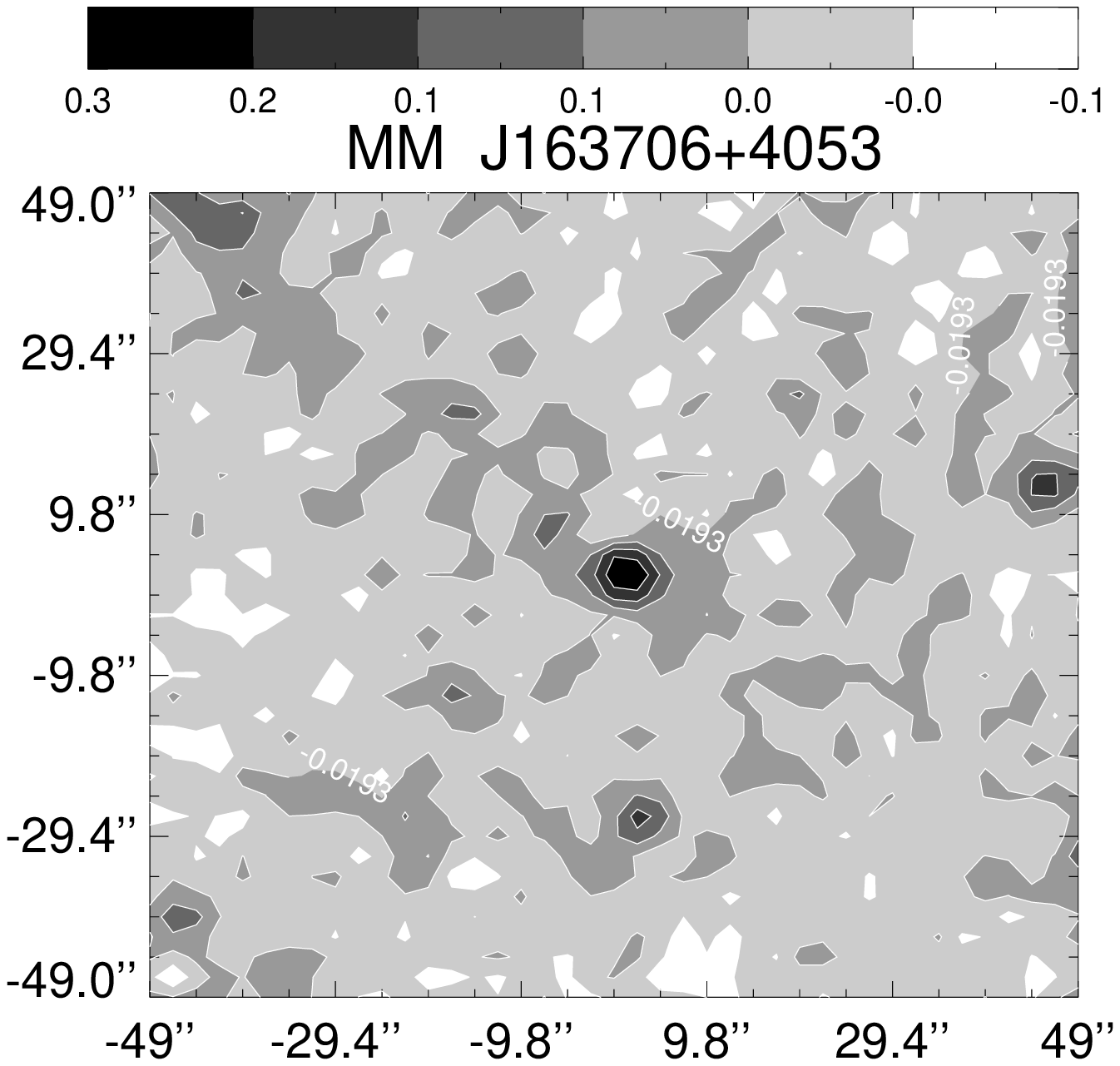,width=6.5cm}}
\vspace{-0.7cm}

\hbox{\hspace{-0.25cm}
\psfig{figure=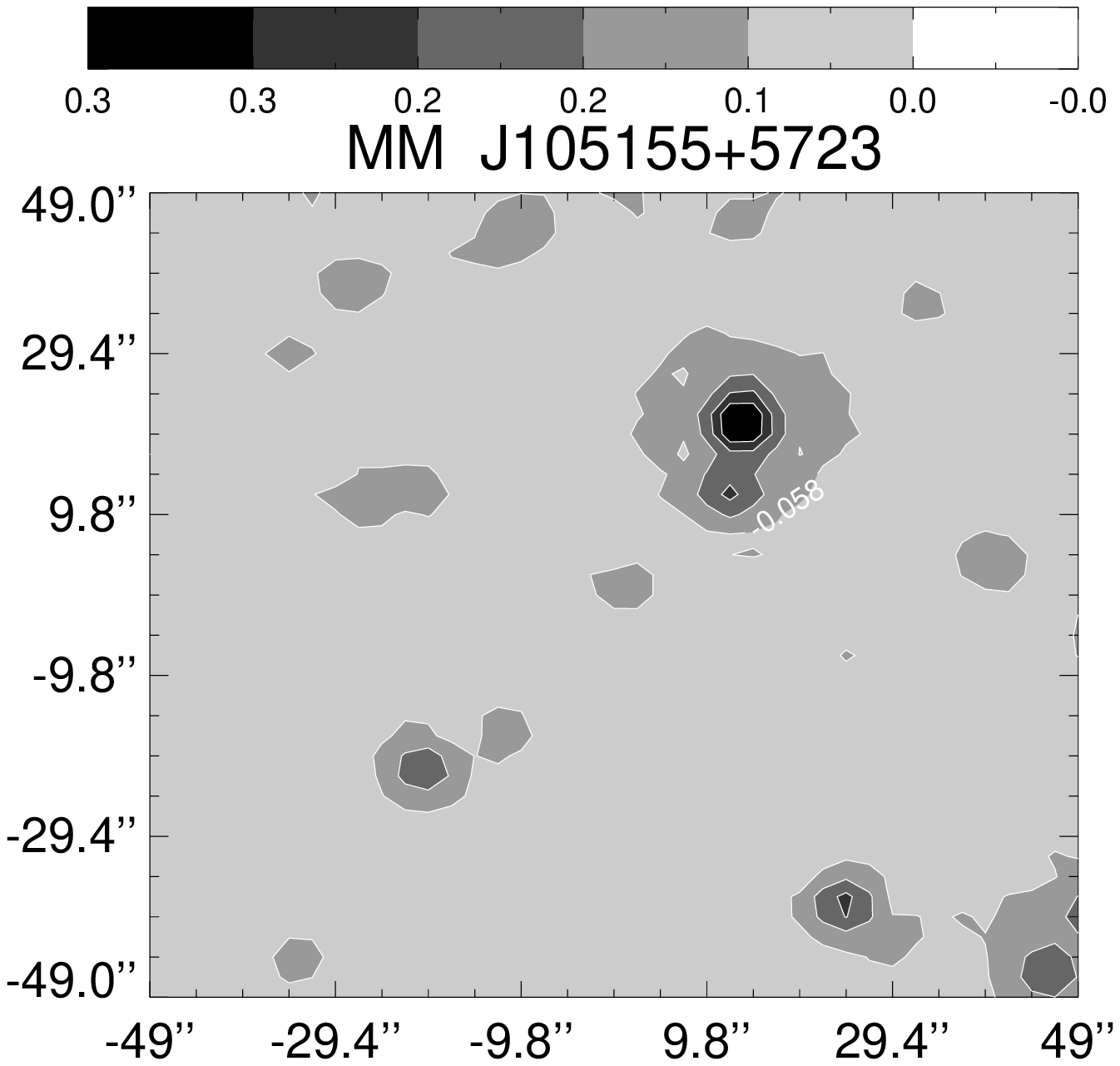,width=6.5cm}
\hspace{-0.7cm}
\psfig{figure=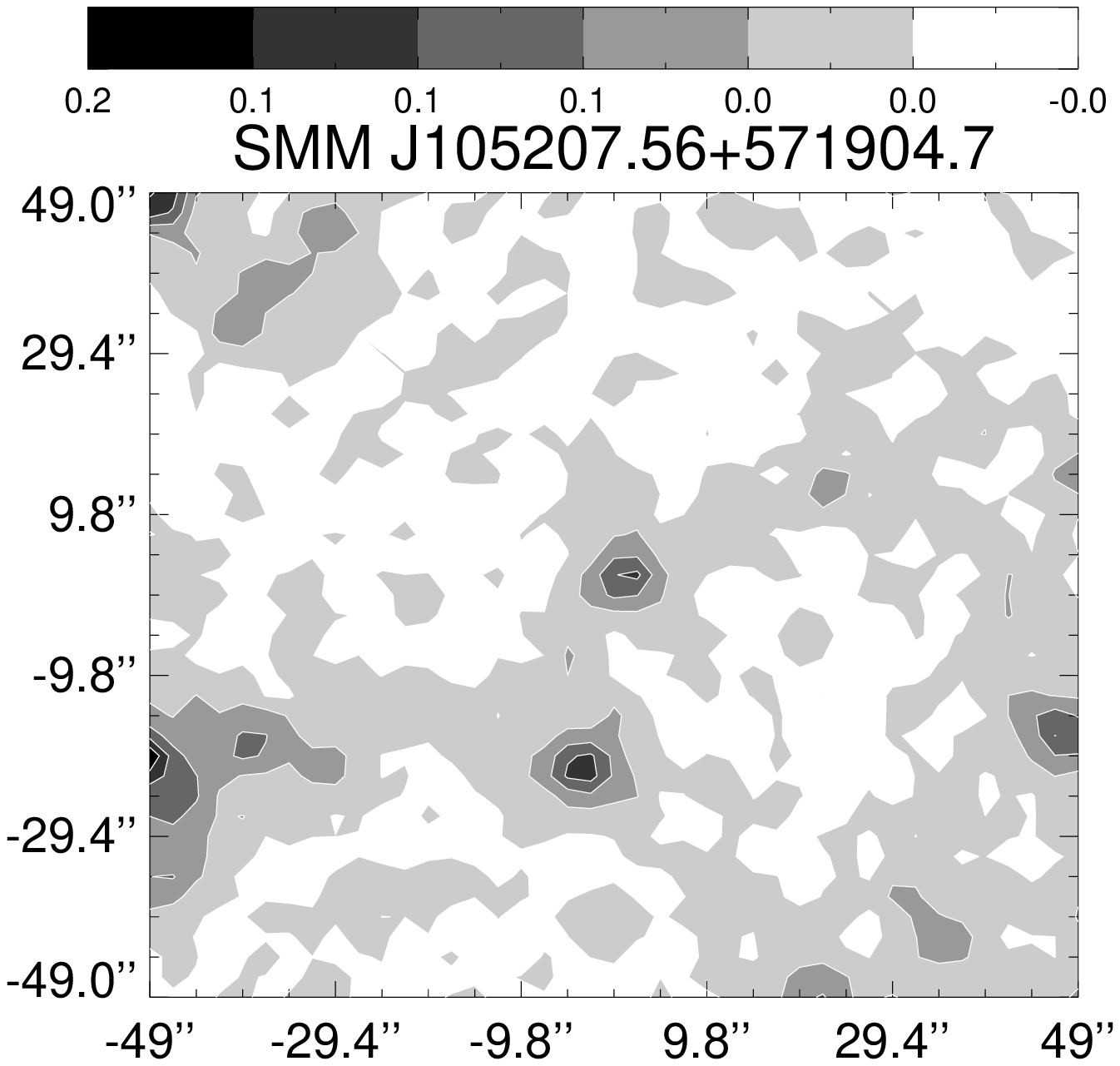,width=6.5cm}
\hspace{-0.7cm}
\psfig{figure=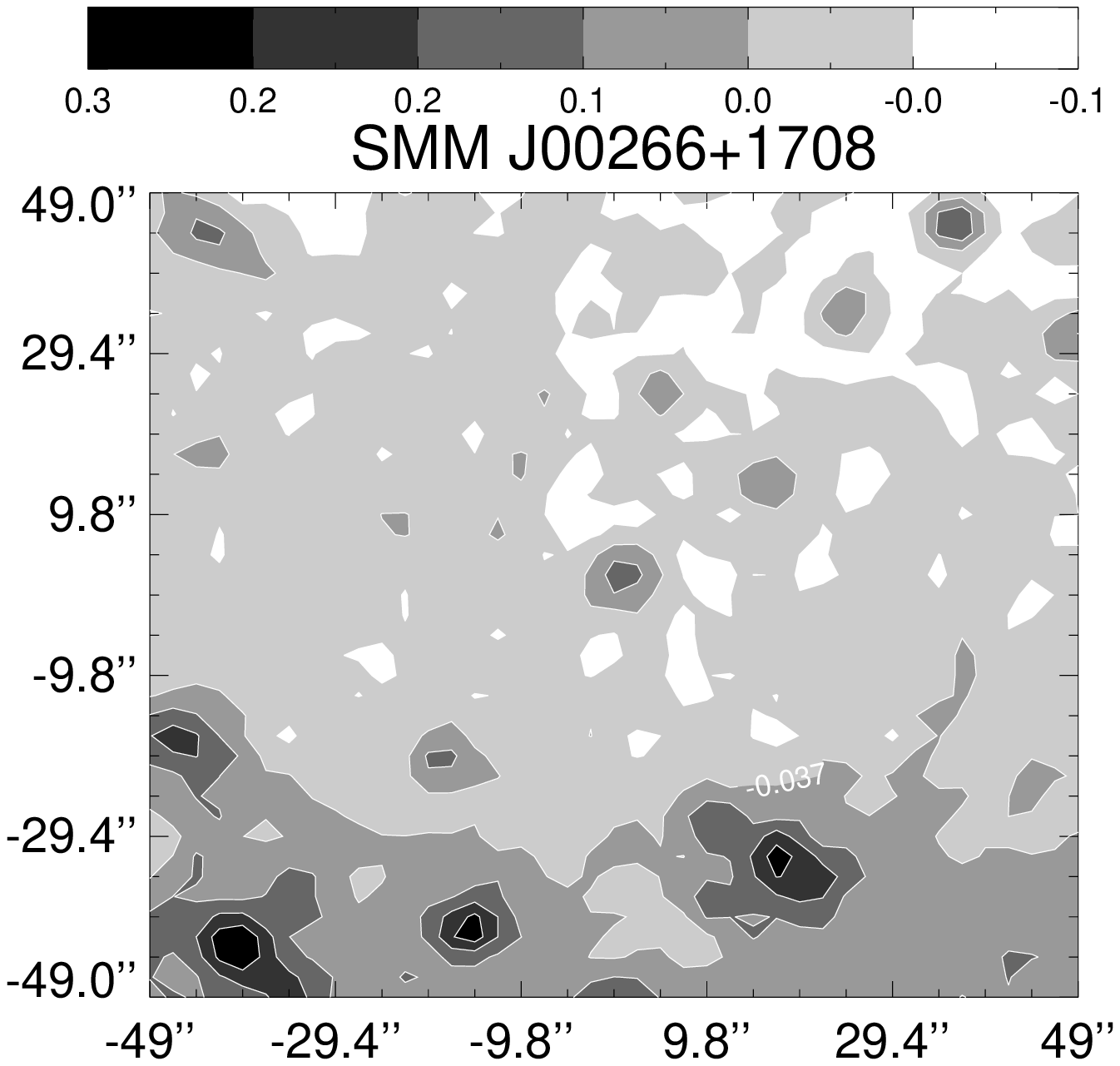,width=6.5cm}}
\vspace{-0.7cm}

\hbox{\hspace{-0.25cm}
\psfig{figure=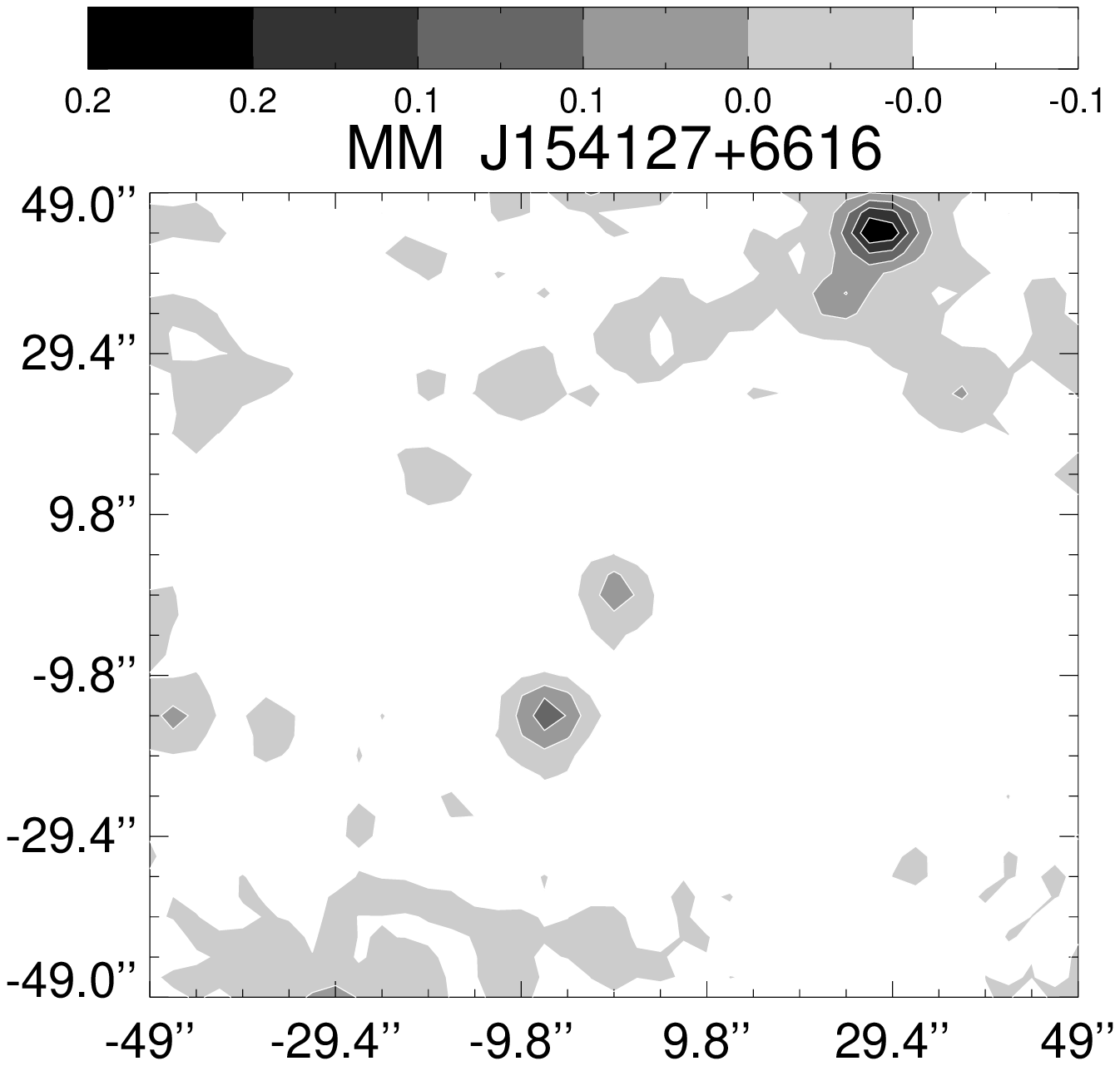,width=6.5cm}
\hspace{-0.7cm}
\psfig{figure=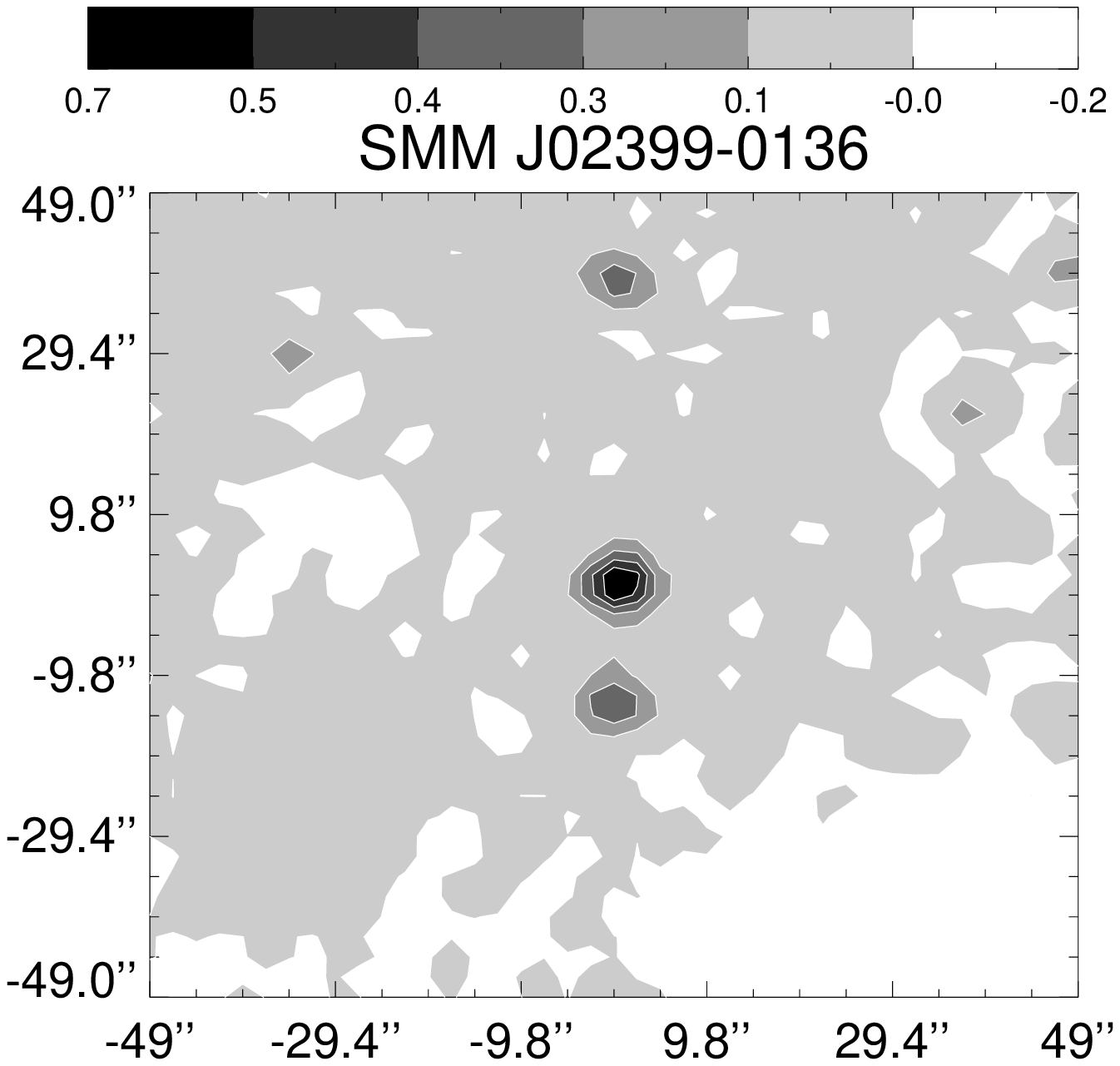,width=6.5cm}
\hspace{-0.7cm}
\psfig{figure=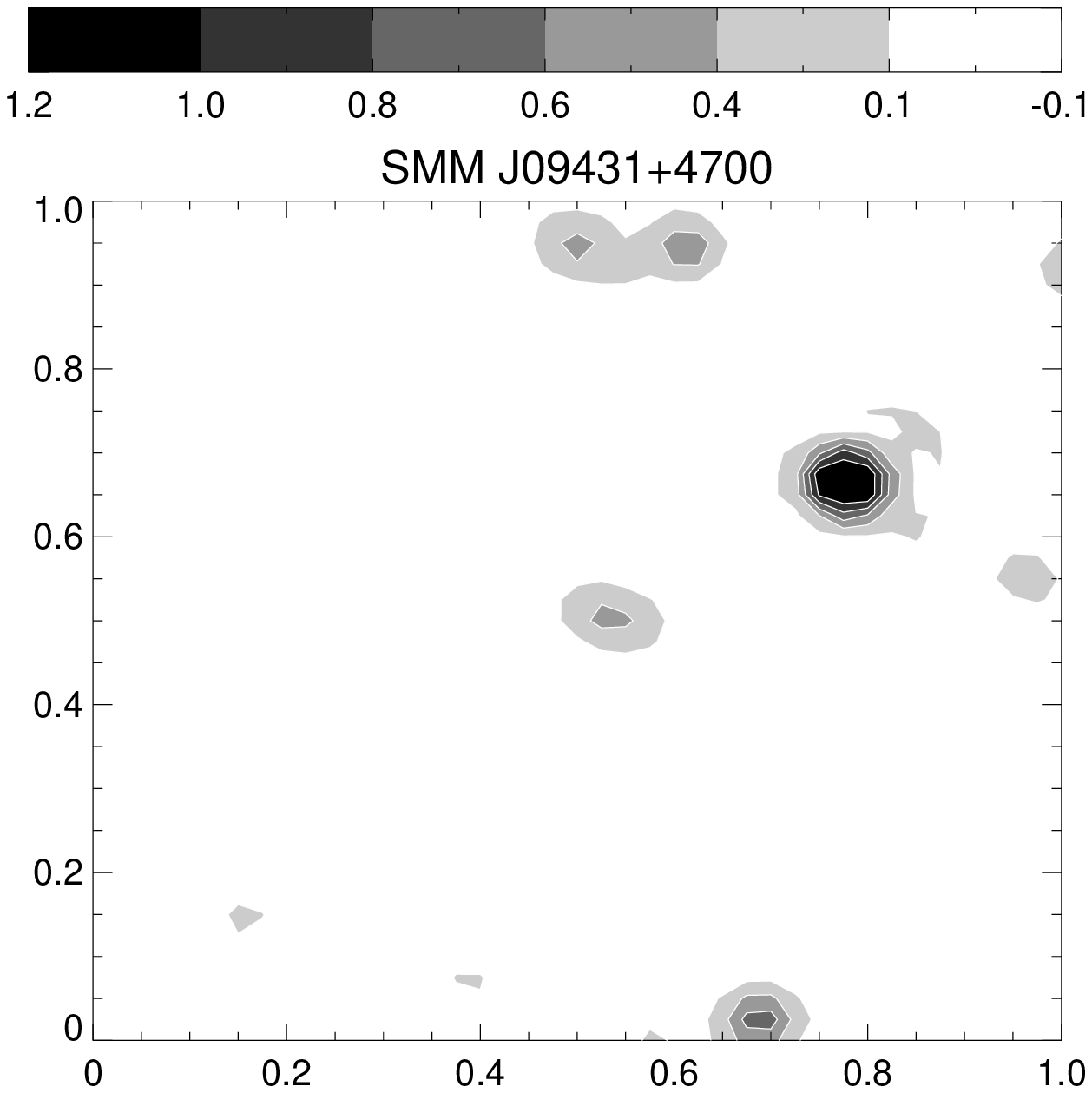,width=6.5cm}}
\vspace{-0.5cm}
\caption{MIPS 24$\mu$m images of the SMGs listed in Table~\ref{obs.tab}.
Images are background subtracted, centered on target and displayed in
a field--of--view of $100'' \times 100''$. Gray scale index in MJy/sr
as indicated and contours at 50\%, 33\%, 25\% and 20\% level from the
brightest pixel in the image. North is up and East is left.
\label{i24.ps} }
\end{figure*}

MIPS raw data are processed by the Spitzer pipeline (version S16.1;
Gordon et al. 2005) to a flux calibrated mosaic image. The various
mosaic images of a particular target and channel and from the
different programs and observers are combined to a final image using
SWARP\footnote{available at: {\tt http://terapix.iap.fr/soft/swarp}}.
For the 70 and 160$\mu$m bands we use the filtered mosaic images of
the pipeline (see MIPS Data Handbook). The final image has higher
redundancy and signal--to--noise than the one obtained from data of a
particular observing sequence.  In the 24$\mu$m band all sources are
detected and final images are shown in Fig.~\ref{i24.ps}. In this band
SMGs appear well separated from other sources.  In the other channels,
at 70 and 160$\mu$m, with the exception of SMM~J02399-0136, SMGs remain
undetected. The flux is derived using an aperture centered on the
first Airy ring and a 2 pixel wide background annulus outside the
second Airy ring; colour and PSF correction factors are applied. The
photometric error is better than 10\%.  MIPS photometry of the SMGs is
given in Table~\ref{obs.tab} and agree with IRS. We verified our
procedure on the calibration standard star HD106252.  For this star
all MIPS 24$\mu$m data are pipeline processed and mosaic images
coadded with SWARP to a final image. On this image we measure a flux
which is consistent to within 2\%  with the flux measured by
Engelbracht et al. (2007) on the same star.

\begin{table}
\caption{Sample of SMGs with redshift and MIPS photometry or 3$\sigma$
upper limits (in mJy). \label{obs.tab}}
\begin{center}
\begin{tabular}{|l | c |c c c|}
\hline
& & & & \\
Name                    & $z$  &24$\mu$m&70$\mu$m& 160$\mu$m \\
& & & & \\
\hline
SMM~J163658.78+405728.1 & 1.2  & 0.45 & $<$4.9 & $<$30 \\
SMM~J030227.73+000653.5 & 1.4  & 0.23 & $<$8.8 & $-$ \\
MM~J163639+4056         & 1.5  & 0.23 & $<$5.5 & $<$21 \\
MM~J163650+4057         & 2.4  & 1.02 & $<$5.5 & $<$22 \\
SMM~J09429+4659         & 2.4  & 0.19 & $<$6.8 & $<$23 \\
MM~J163706+4053         & 2.4  & 0.41 & $<$5.1 & $<$17 \\
MM~J105155+5723         & 2.7  & 0.18 & $<$2.2 & $<$16 \\
SMM~J105207.56+571904.7 & 2.7  & 0.24 & $<$2.0 & $<$13 \\
SMM~J00266+1708         & 2.7  & 0.32 & $-$    & $-$ \\
MM~J154127+6616         & 2.8  & 0.24 & $-$    & $-$ \\
SMM~J02399-0136         & 2.8  & 1.24 & $15^{\dagger}$ &  16 \\
SMM~J09431+4700         & 3.4  & 1.04 & $<$7.2 & $<$22 \\
\hline
\end{tabular}
~~~~~~~~~~~~~~~~~~~~~~~~~~~~~~~~~~~~~~~~~~~~~~~~~~$^{\dagger}$ {\scriptsize Tentative 
detection:~brightest source pixel appears $\sim 8''$ off from NED position.}
\end{center}
\end{table}

\begin{figure*}
\centerline{\psfig{figure=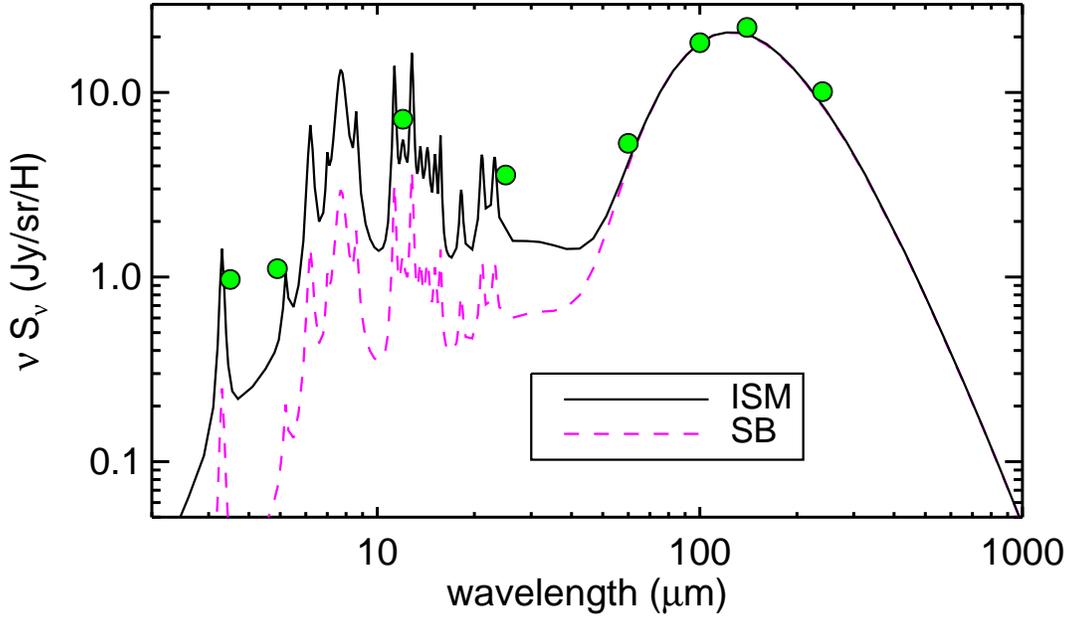,angle=0,width=15.cm}}
\caption{Comparison of cirrus models with DIRBE observations for High
 Galactic Latitudes (Arendt et al. 1998). The black solid line shows
 the emission from dust illuminated by the Mathis et al. (1983)
 interstellar radiation field and a $\psi$ of 1. The dust model is
 described in Sect.~3. The magenta dashed line shows the emission from
 a dust mixture more typical for a starburst where the PAH abundance
 is reduced by a factor of 5.}
\label{isrf.ps}
\end{figure*}

\begin{figure*}
\centerline{\psfig{figure=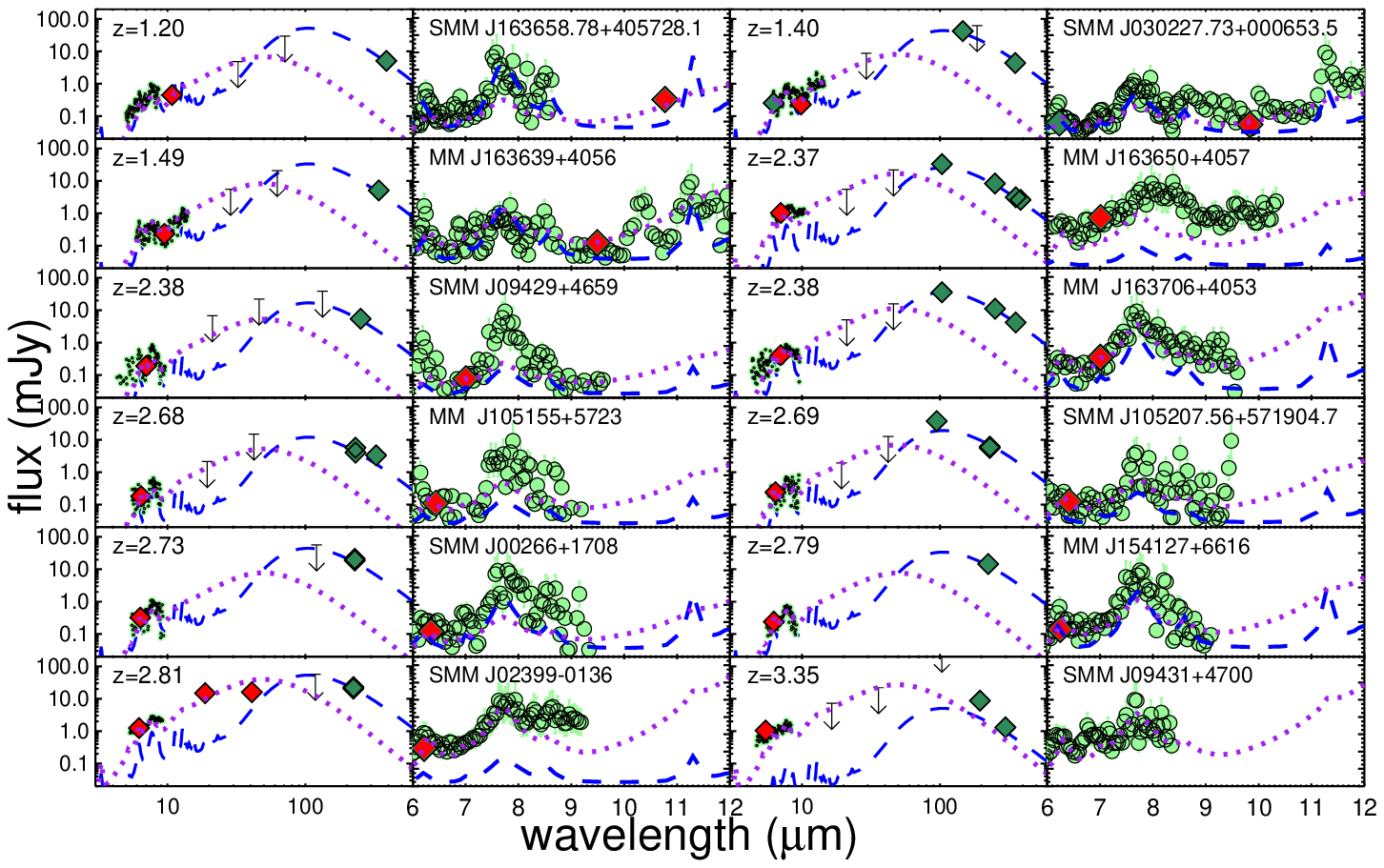,angle=0,width=15.cm}}

\caption{Comparison of the spectra of SMGs with pure cirrus (blue
dashed) and pure starburst (purple dotted) models.  The cirrus model
assumes that the stellar population that illuminates the dust is
250Myrs old and has a bolometric intensity which is $\psi = 5$ times
that of the interstellar radiation field in the solar neighborhood.
The starburst model assumes a nuclear radius of 3kpc, a visual
extinction of 18mag and a luminosity of $L=10^{13.1}$\Lsun but scaled
to fit the mid-infrared photometry. Data from {\it this work} and
Smail et al. (2002), Scott et al. (2002), Webb et al. (2003), Ivison et
al. (2005), Greve et al. (2005), Scott et al. (2006),
Men\'{e}ndez-Delmestre et al. (2006), Kov\'{a}cs et al.  (2006),
Valiante et al. (2007) and Zemcov et al. (2007).  For each galaxy two
panels are shown: one covering the $3 - 600 \mu$m and the other the $6
- 12 \mu$m rest--frame wavelength range; the vertical scale of the $6
- 12 \mu$m panel is linear between $0 - 1.5$mJy. \label{cirrus.ps}}

\end{figure*}

\section{Cirrus and starburst-only models}

 We first compare the data of the galaxies in our sample with pure
cirrus and pure starburst models. For the cirrus emission we follow an
approach similar to that of ERR03 except that we do not attempt to
link the optical-UV emission with the infrared in a self-consistent
manner. We first determine the near-infrared to UV spectrum by
assuming an age for the galaxy and a star formation history. We use
the stellar population synthesis model of Bruzual \& Charlot (1993)
with a Salpeter IMF and stellar masses in the range 0.1-125$\
M_\odot$. For this study we assume, as in ERR03, an age of 250Myrs and
a star formation rate that is constant with time.  For the last 5Myrs
we assume that the stars are embedded in the molecular clouds in which
they formed so they do not contribute to the starlight that is
illuminating the cirrus dust. We then use the parameter $\psi $ to
scale the spectrum emitted by the stars; $\psi $ is defined to be the
ratio of the bolometric intensity of starlight to the bolometric
intensity of the interstellar radiation field in the solar
neighborhood (Mathis et al. 1983).  In Fig.~\ref{isrf.ps} we compare
the spectrum emitted by dust that is illuminated by the Mathis et al.
interstellar radiation field for the solar neighborhood ($\psi = 1$)
with DIRBE data for High Galactic Latitudes (Arendt et al. 1998). To
determine the spectrum emitted by the dust we use the latest dust
model by Siebenmorgen et al. (2001). It consists of silicate and
amorphous carbon grains with a size distribution: $n(a) \propto
a^{-3.5}$, $300 {\rm \AA} \leq a \leq 2400 \rm {\AA}$, a population of
small graphite grains with $n(a) \propto a^{-4}$, $10\rm{\AA} \leq a
\leq 80 \rm{\AA}$ and PAHs. There are small PAHs with $N_C = 50$ and
$N_H = 12$ and large PAHs with $N_C = 300$ and $N_H = 48$, where $N_C$
is the number of C atoms and $N_H$ the number of H atoms of a PAH.  By
mass 63\% of the dust is in silicates and 37\% in carbon of which 80\%
is amorphous, 10\% graphitic, 5\% in small and 5\% in large PAHs.  The
element abundance, with respect to H, of C and Si in grains is 200ppm
and 31ppm, respectively.  The gas--to--dust mass ratio, which allows
conversion from dust into gas mass, is 125. Such a dust mixture
produces a reddening in rough agreement with the standard interstellar
extinction curve for $R_V = 3.1$ and can be considered to be typical
for interstellar dust in a quiescent galaxy like our own. However, in
the more active environment of a starburst galaxy some of the small
grains and PAHs are destroyed by the harder photon and stronger
radiation field (e.g. Omont 1986, Leach et al. 1989a,b, Voit 1991,
Voit 1992, Rapacioli et al. 2006, Siebenmorgen \& Kr\"ugel 2009), thus
reducing their abundance. Vega et al. (2005) even suggest that
all PAHs are destroyed in the molecular clouds that constitute
the starburst and that the PAH emission of infrared galaxies
arises from cirrus dust. The predicted spectrum of the dust emission
that is illuminated by the Mathis et al. interstellar radiation field
but with reduced small grain abundances is shown in Fig.\ref{isrf.ps}.
We consider that for a starburst 5\% of carbon in the grains is in
small graphites and 2\% of C is in PAHs. For the rest of the paper we
will assume these parameters of the dust for our cirrus and starburst
model.

In Fig.~\ref{cirrus.ps} we compare our cirrus model with the data of
the galaxies in our sample. Clearly we cannot constrain the value of
$\psi $ with the presently available far-infrared and submillimeter
data. However, in order to be consistent with the finding of Coppin et
al. (2008) that most  SMGs emit at a temperature of 28K we use $\psi =
5$. It is important to note that the estimate of Coppin et al. is
based on fitting a modified blackbody to the 350$\mu m$ and 850$\mu m$
data whereas our cirrus model considers a distribution of grain species
and sizes each with its own temperature. 
 After normalizing the cirrus model to
the 850$\mu m$ data point we find that for a couple of the objects
(SMM~J163658.78+405728.1 at a redshift of 1.2 and MM~J163639+4056 at a
redshift of 1.5) the cirrus model seems to be adequate for explaining
the spectral energy distribution. For SMM~J10521+5719 a cirrus model
with $\psi =18$ is also adequate for explaining the SED. For the rest
of the objects the model falls short of matching the observed
mid-infrared spectrum suggesting either that the PAH abundance is
higher than assumed or there is contribution in the mid-infrared from
a starburst. In Sect.~4 and 5 we will explore the latter possibility.

In Fig.~\ref{cirrus.ps} we also explore the possibility that a
starburst alone can explain the complete spectral energy
distribution. To do that we take a single starburst spectrum from the
SED library of Siebenmorgen \& Kr\"{u}gel (2007). We choose a
starburst model with a total luminosity of $10^{13.1}$\Lsun, a nuclear
radius of 3kpc and a visual extinction from the edge to the center of
18mag. For each galaxy this model spectrum is normalized to the
mid-infrared observations. It is clear that a starburst-only model
falls short of matching the submillimeter photometry by at least an
order of magnitude. The model is described in Sect.~5 but it is
important to note here that as the starburst radius increases the dust
gets colder and the spectrum shifts to longer wavelengths. Starburst
models with a radius of 10kpc come close to matching the complete
spectral energy distribution but as we discuss in Sect.~6 such
extended starbursts are excluded by millimetre interferometry.

\begin{figure*} 
\centerline{\psfig{figure=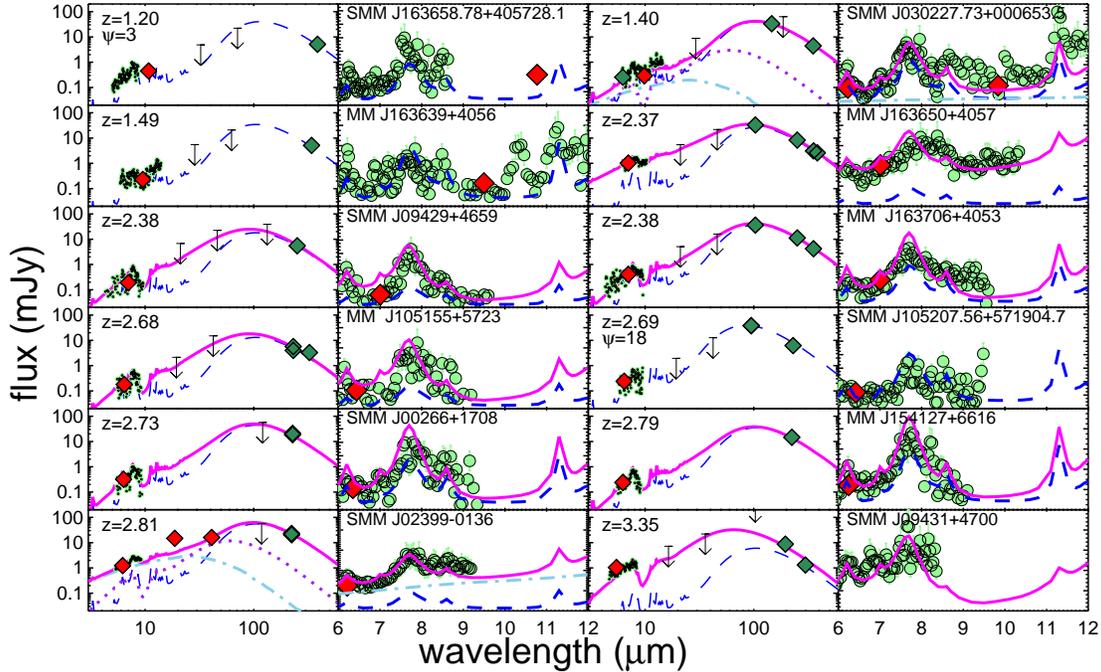,angle=0,width=15.cm}}
\caption{Comparison of the spectral energy distributions of the SMGs
with a combination of evolutionary starburst, cirrus and AGN torus
models.  Unless otherwise indicated the cirrus emission is computed
with a $\psi =5 $ scaling of the interstellar radiation field. The
data and layout of the figure is as described in
Fig.\ref{cirrus.ps}. Cirrus (blue
dashed), AGN torus (light blue dash--dotted) and total emission
spectrum (magenta solid) is indicated.  Derived quantities are given
in Table~{\ref{mod.tab}}. \label{ane.ps}}
\end{figure*}

\section{Evolutionary starburst models}

Efstathiou, Rowan-Robinson \& Siebenmorgen (2000; hereafter ERRS)
presented a starburst model that combined a simple model for the
evolution of giant molecular clouds, the stellar population synthesis
model of Bruzual \& Charlot (1993) and detailed radiative transfer
that included the effect of temperature fluctuating small grains with
dust particle radius $a < 100$\AA \/ and PAHs to account for the
detected infrared emission bands (Siebenmorgen \& Kr\"{u}gel 1992). An
important feature of the ERRS model is that the expansion of the HII
region leads to the formation of a narrow shell of gas and dust.  This
naturally explains the fact that the near- and mid-infrared spectra of
starburst galaxies are not dominated by emission from hot large
($a>100$\AA) dust grains but by the PAHs emission. As in ERRS we
assume that the initial $A_V$ of the molecular clouds that constitute
the starburst is $50$mag but we use the grain model described in Sect.~3. We
further assume a constant star formation rate and an age of 5Myr. The
assumed starburst age is very poorly constrained by our modeling
but as we discuss in section 6 a value of 5Myr can explain the fact
that the inferred luinosities of the starburst and cirrus components
are comparable.

In Fig.~\ref{ane.ps} we present fits to the galaxies in our sample
with a combination of starburst and cirrus. We first normalize the
cirrus model at 850$\mu m$ and then scale the starburst model so that
the combination of starburst and cirrus gives the best fit to the
mid-infrared spectroscopy and is consistent with the far-infrared
photometry. As discussed in Sect.~3, for three of the objects we do
not need any starburst contribution for explaining the SED. For two
of the objects we also find evidence for an AGN component which we
model with the tapered discs of Efstathiou \& Rowan-Robinson
(1995). The more luminous of the two objects (SMM~J02399-0136), has
been classified as a Seyfert 2 by Smail et al. (2002). SMM~J02399-0136
is also the object that has been detected at 70 and 160$\mu m$. The
luminosities of the three components and the associated dust masses
are given in Table~2. In the ERRS model an increase in
the luminosity translates into an increase in the number of molecular
clouds that constitute the starburst and therefore the dust mass.

\begin{figure*}
\centerline{\psfig{figure=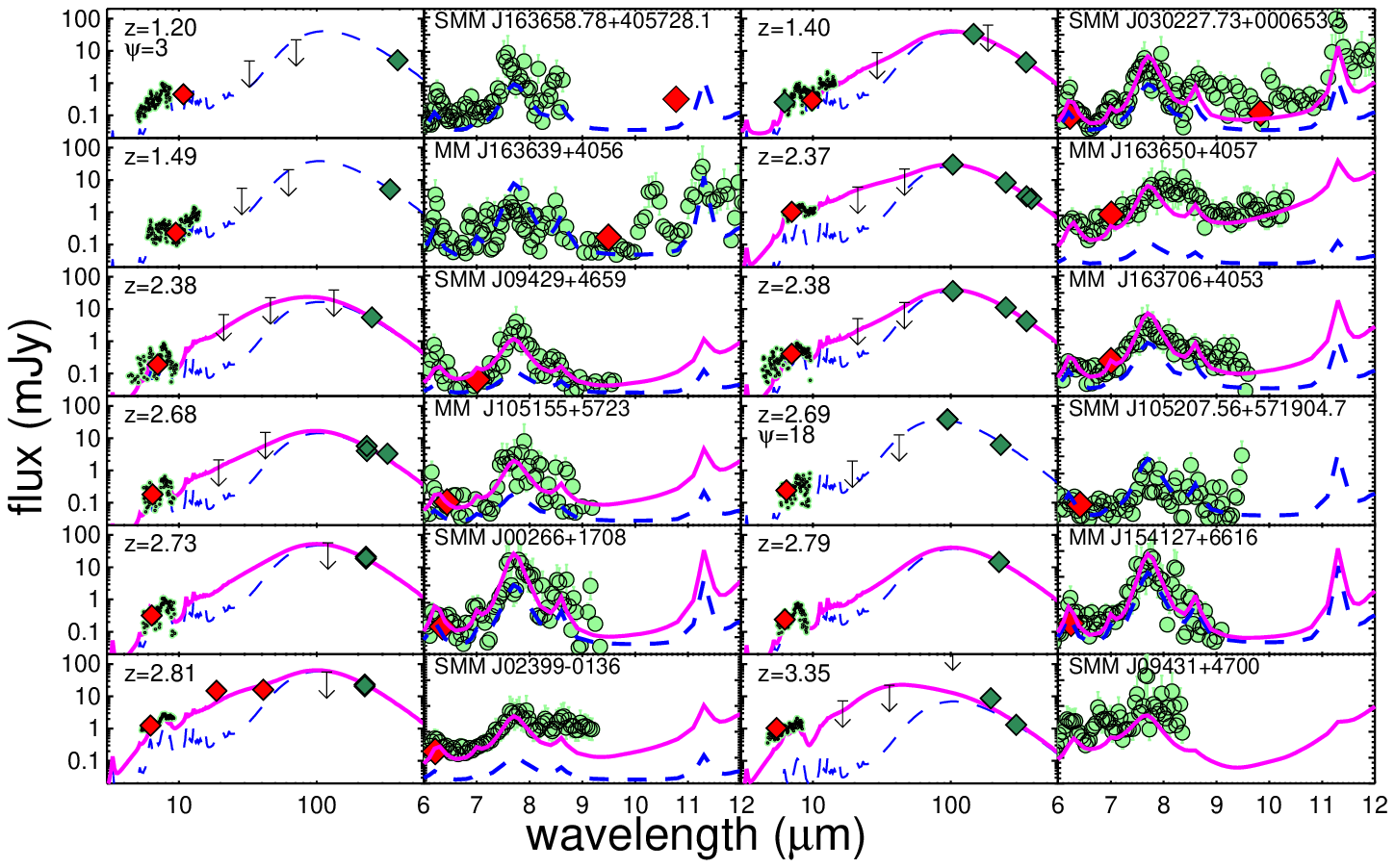,angle=0,width=15.cm}}
\caption{Comparison of the spectral energy distributions of the SMGs
with a combination of hot spot starburst and cirrus models.  Unless
otherwise indicated all cirrus models assume a $\psi =5 $ scaling of
the interstellar radiation field. Data and layout of the figure is as
in Fig.~\ref{cirrus.ps}.  Cirrus (blue dashed) and total emission
spectrum (magenta solid) is indicated.  Derived quantities are given
in Table~{\ref{mod.tab}}. \label{rs.ps}}
\end{figure*}

\section{Hot spot starburst models}

Siebenmorgen \& Kr\"ugel (2007) presented a starburst model which is
an evolution of an earlier model described by Kr\"ugel \& Tutukov
(1978) and Kr\"ugel \& Siebenmorgen (1994).  The model assumes that
the stars are divided in two classes: OB stars that are surrounded by
dense clouds and constitute so-called hot spots and other stars (old
bulge stars or massive stars) that are dispersed in the diffuse medium (see below).
The hot spots determine the mid infrared part of the emission
spectrum. The outer radius of the hot spots is determined by the
condition of equal heating of the dust by the stars and the ambient
interstellar radiation field. Both classes of stars are represented in
the equation of radiative transfer by continuously distributed source
terms. It is assumed that the number density of the hot spots and of
the other stars falls off with the radius of the starburst as
$r^{-1.5}$.

In addition to the dust in the hot spots the model assumes that the
volume of the starburst is filled by dust which is uniformly
distributed and gives rise to a total extinction $A_V^{'}$ from the
outer radius $R$ of the galactic nucleus to its center. In this
homogeneous density model the parameter $A_V^{'}$ is directly related
to the dust mass $M_{SB}^{'}$ and only one of them is independent. The
other model parameters are the total luminosity $L_{SB}^{'}$ and the
starburst radius $R$.  The OB stars are assumed to be confined to the
central 350pc whereas the bulge stars fill the whole volume.

In Fig.~\ref{rs.ps} we combine starburst models computed with the
method of Siebenmorgen \& Kr\"{u}gel (2007) with cirrus models and compare
them with the data of the objects in our sample.  As in the case of
the evolutionary models we first normalize the cirrus model at
850$\mu m$. Then we search in the SED library to find a starburst
model which, after scaling to the distance of the object but without
further normalization, best fit the mid-infrared spectroscopy. The
starburst radius is fixed at 3kpc to be consistent with the sizes
inferred from interferometry for these galaxies (Sect.~6).  Unless
otherwise indicated the value of $\psi $ is assumed to be 5 as before.

\begin{table*}
\caption{Derived parameters for each galaxy. \label{mod.tab}}
\begin{center}
\begin{tabular}{|l|cccc|lcl|}
\hline
   & &&&& && \\
Name    &  \multicolumn{4}{|c|}{Evolutionary models} & 
\multicolumn{3}{|c|} {Hot spot models} \\
   & &&&& && \\
   & $\log L_{\rm{SB}}$ & $\log L_{\rm{C}}$  
& $\log L_{\rm{AGN}}$  & $M_{\rm {C}}$ 
&  $\log L_{\rm{SB}}'$  &  $\log L_{\rm {C}}'$  & $A_{\rm {V}}$    \\
          &          \scriptsize{[\Lsun ]}             & \scriptsize{[\Lsun ]}           
& \scriptsize{[\Lsun ]}     & \scriptsize{[$10^9$\Msun ]}          &   \scriptsize{[\Lsun ]}     
&  \scriptsize{[\Lsun ]}       & \scriptsize{[mag]}         \\
& (1) & (2)  & (3)  & (4) & (5)  &  (6)  & (7)   \\
  & &&&& && \\
\hline
{\scriptsize SMM~J163658.78+405728.1}  &-     &12.2 &-     & 1.3 &-           &12.2&-       \\
{\scriptsize SMM~J030227.73+000653.5}  & 11.7 &12.2 &11.3  & 1.5 & 12.0       &12.3&7       \\
{\scriptsize MM~J163639+4056      }    &-     &12.3 &-     & 1.4 &-           &12.3&-       \\
{\scriptsize MM~J163650+4057     }     & 12.8 &12.4 &-     & 2.2 & 12.7       &12.6&9       \\
{\scriptsize SMM~J09429+4659    }      & 12.5 &12.3 &-     & 1.5 & 12.5       &12.4&36      \\
{\scriptsize MM~J163706+4053   }       & 12.4 &12.6 &-     & 3.0 & 12.4       &12.7&18      \\
{\scriptsize MM~J105155+5723  }        & 12.5 &12.2 &-     & 1.3 & 12.3       &12.4&18      \\
{\scriptsize SMM~J105207.56+571904.7}  &-     &12.9 &-     & 3.6 &-           &12.9&-       \\
{\scriptsize SMM~J00266+1708 }         & 12.5 &12.8 &-     & 4.5 & 12.6       &12.9&36      \\
{\scriptsize MM~J154127+6616 }         & 12.4 &12.7 &-     & 3.7 & 12.3       &12.8&36      \\
{\scriptsize SMM~J02399-0136   }       & 12.7 &12.9 &12.7  & 6.0 & 13.0       &13.0&18      \\
{\scriptsize SMM~J09431+4700 }         & 13.2 &12.0 &-     & 0.9 & 13.2       &12.3&18      \\
\hline
\end{tabular}
\end{center}

 {\scriptsize Column (1): starburst luminosity of the evolutionary model.}
 {\scriptsize Column (2): cirrus luminosity of the evolutionary model.}
 {\scriptsize Column (3): AGN torus luminosity.}
 {\scriptsize Column (4): dust mass of the cirrus component.}
 {\scriptsize Column(5): starburst luminosity of the hot spot model.}
 {\scriptsize Column (6): cirrus luminosity of the hot spot model.}
 {\scriptsize Column (7): visual extinction of the hot spot model measured from the surface to the center of
the galaxy.} 

\end{table*}

\section{Discussion and Conclusions}

Since the original suggestion of ERR03 and Kaviani et al. (2003) that
SMGs are colder and more extended than local ULIRGs a number of studies
have given support to this idea. Rowan-Robinson et al. (2004)
discovered a number of luminous cirrus galaxies in the ELAIS
survey. Chapman et al. (2004) found that the majority of SMGs in their
sample are more extended in the radio than local ULIRGs. This finding
was confirmed by Biggs \& Ivison (2008) who found a median size of
~5kpc for the SMGs in their sample.  However, Tacconi et al. (2006)
showed that the SMGs in their sample cannot be more extended than $\sim
4$kpc.

The evolutionary starburst model of ERRS does not make a prediction
about the size of the starburst but a lower limit can be obtained by
calculating the minimum volume needed to contain the molecular clouds
that constitute it. As noted by ERRS a molecular cloud illuminated by
a 10Myr old instantaneous burst has a luminosity of $\sim 3 \times
10^8 \ \rm{L}_\odot$ and a radius of 50pc. The minimum radius $R$ of the
sphere that is needed to contain a starburst of luminosity $L$ is
therefore $R = 0.05 ( {{L}/{3\times 10^8 \ \rm{L}_\odot}} )^{1/3}$
kpc.  So for a ULIRG with a luminosity of $3 \times 10^{12} \
\rm{L}_\odot$, the radius is $R \sim 1$kpc.  A plausible geometry for
SMGs is therefore a region of $2-4$kpc radius
filled with diffuse dust and in which an ensemble of molecular clouds
is embedded.

Kov\'{a}cs et al. (2006) reported 350$\mu$m observations of SMGs which
suggest that they are colder (mean temperature $\sim 35$K) than local
ULIRGs. This was confirmed by Coppin et al. (2008) who reported
350$\mu$m observations of SMGs found in the SHADES survey. In fact
Coppin et al. find that most SMGs in their sample are actually even
colder (28K).
  
Clements et al. (2008) presented spectral energy distributions of SMGs
detected in the SHADES survey and for which there are also
supplementary optical and {\it Spitzer} data. They found that out of
the 33 sources in their sample 8 can be fitted by a cirrus template
whereas most of the other sources can be fitted by an Arp220
template. The fraction of cirrus dominated galaxies is therefore in
good agreement with our estimate. The problem with the Arp220
interpretation for the other objects, however, is that none of the
objects with mid-infrared spectroscopy from {\it Spitzer} show the
deep silicate absorption features characteristic of the Arp220
template.
  
Pope et al. (2008) presented mid-infrared spectroscopy from Spitzer for
a sample of 13 sources in the GOODS field. They find that the
spectroscopy cannot be fitted with an Arp200-type template but
instead the average spectrum can be fitted by an M82-type one.
  
Farrah et al. (2008) presented mid-infrared spectra of high-redshift
ULIRGs in a very narrow redshift range of $1.71 \pm 0.15$.  The ULIRGs
in the sample of Farrah et al. show mid-infrared spectra similar to
local starburst galaxies which are two or more orders of magnitude
less luminous in the infrared. Farrah et al. suggest that one of the
most likely interpretations of their result is that star formation in
the ULIRGs of their sample is extended on scales of 1-4kpc which is
similar to the sizes of SMGs observed with radio or
millimeter interferometry.

The median dust mass we derive for the cirrus component is $2.2 \times
10^9 \ \rm{M}_\odot$. For comparison Coppin et
al. find a median value of $0.9 \times 10^9 \ \rm{M}_\odot$ but note
that most of the difference can be attributed to a factor of three
difference in the assumed absorption coefficient in the
submillimeter. As can be seen from Table~{\ref{mod.tab}} the
luminosities in the starburst and cirrus components are comparable.

The picture we are proposing for SMGs is the following: We assume that
stars have been forming continuously in the galaxy for the last
250Myr. Our assumption about the age of the galaxies is in agreement
with the estimate of Tacconi et al. (2008).  We further assume that
the stars that formed in the last 5Myr are still embedded in the
molecular clouds in which they formed. Using Bruzual \& Charlot (1993)
models with a star formation rate of $1000\ \rm{M}_\odot$/yr we find that
the bolometric luminosity of the 5Myr old starburst is $3.2 \times
10^{12} \ \rm{L}_\odot$ whereas the luminosity of the rest of the stars,
which are between 5 - 250Myr old, is $3.5 \times 10^{12} \ \rm{L}_\odot$.  Assuming that
the optical/UV radiation emitted by the 5-250Myrs old stellar
population is obscured by dust with an $A_V$ of 1mag, and that all the
energy absorbed in the optical/UV is re-radiated at a temperature of
28K, this gives a cold dust luminosity of $2.5 \times 10^{12} \
\rm{L}_\odot$. So this explains the fact that the derived luminosities
of the starburst and cirrus components are comparable. 
 The median of the log of luminosities of the cirrus
component of the 12 galaxies (in solar units)  is 12.4 for the evolutionary models and
12.6 for the hot spot models. The star formation rate for the typical
galaxy is therefore $\sim 1000-1600 \ \rm{M}_\odot$/yr which is in
good agreement with the estimates of Ivison et al. (2002).  It is
interesting to note that the timescale we infer for the {\it
{molecular cloud}} phase of star formation is similar to the value
found by Granato et al. (2000) and Efstathiou \& Rowan-Robinson (2003)
for normal quiescently star-forming galaxies.

 The analysis presented in this paper is clearly limited by the lack of
data in the rest-frame far-infrared  where both the starburst and
cirrus components peak. Measurements with {\em Herschel}, whose launch
is now imminent, will allow us to test the model outlined above.
They will allow us in particular to put a stronger constraint on the
intensity of starlight in the case of the cirrus component and possibly
differentiate between the two starburst models considered in this paper.

\section*{Acknowledgments}

We thank Kar\'{i}n Men\'{e}ndez-Delmestre and Elisabeta Valiante for
providing the spectra of the SMGs in electronic
form, Andreas Seifahrt for introduction to SWARP and Helmut
Dannerbauer for his comments on an earlier draft of this work.

\end{document}